\documentstyle[preprint,revtex,eqsecnum]{aps}
\tightenlines
\begin{document}
\draft
\preprint{}
\begin{title}
Correlation effects on the band gap of conducting polymers
\end{title}
\author{C.Q.\ Wu$^*$}
\begin{instit}
Max-Planck-Institut f\"ur Festk\"orperforschung, 7000 Stuttgart 80\\
Federal Republic of Germany
\end{instit}

\centerline{\it (Received 31 August 1992)}

\begin{abstract}
By applying the projection technique to the computation of excitation
energies, we study the correlation effects on the band gap of
conducting polymers. In the presence of an additional electron or
hole, the correlation induces a polarization cloud around the
additional particle, which forms a polaron. For the excitation
energy of a polaron, there is a competition between a {\it loss} of
the correlation energy in the ground state and a {\it gain} of
polarization energy. For the Hubbard interaction, the {\it loss} of
correlation energy is dominant and correlations increase the band
gap. However, for long-range interactions, the {\it gain} of
polarization energy is dominant and correlations decrease
the band gap. Screening the long-range interaction suppresses
the {\it gain} of the polarization energy so that correlations
again increase the band gap. A small dimerization is always
favorable to the correlation effects. For {\it trans}-polyacetylene,
we obtain the on-site repulsion $U=4.4$ eV and the nearest-neighbor
interaction $V=0.8$ eV. The screening of $\pi$ electrons due to
the polarizability of $\sigma$ electrons is quite strong.
\end{abstract}


\narrowtext
\section{INTRODUCTION}
In conducting polymers, such as {\it trans}-polyacetylene, the
importance of electron-electron interactions has been widely
accepted. Many physical phenomena, for example, the nonvanishing
negative spin density on alternate carbon atoms \cite{htea}, the
relative ordering between states $2^1A_g$ and $1^1B_u$ \cite{bhbk},
the optical absorption associated with neutral soliton \cite{joea},
etc., should be interpreted by coping with electron-electron
interactions, or more precisely, with electron correlations. But
how strong are the correlations? This question has been argued for
many years. A number of theories have been developed over the years
to treat electron correlation in polyacetylene,
among them are the mean field and perturbation theories
\cite{ksmg,skdh,wksk}, the valence bond analysis \cite{zssr},
the Monte Carlo calculation \cite{jh,dktdsm}, Local Ansatz
method \cite{ph,dbkm}, and the correlated-basis-function theory
\cite{wsn,xsea}. Comparison of these results with experiments has
led to conflicting claims about the strength of electron
correlations.

The band gap is an important physical quantity which
is strongly dependent on electron correlations. It can
be determined by the optical absorption and for
{\it trans}-polyacetylene, a value of 1.8 eV is found
\cite{nsea}. It is well known that calculated band gaps come
out much too large when the independent-electron or
self-consistent field (SCF) approximation is made.

Recently, K\"onig and Stollhoff \cite{gkgs} performed an {\it ab
initio} calculation for the ground state of polyacetylene. By
fitting the dependence of the total energy on the dimerization,
they were able to determine the on-site electron-electron interaction
$U=11.5$ eV as well as the effective interaction of electrons on
nearest neighbor sites, $V=2.4$ eV. When the energy gap is
computed within SCF approximation a value of 6.9 eV is found, which
is much larger than the observed one of 1.8 eV. The difference
between these two values is apparently due to correlations.

The projection technique \cite{pf}, developed recently by Fulde
and his coworkers \cite{bf1,bb}, has been successfully applied
to both weakly and strongly correlated electron systems
\cite{bf2,bwf}. In this paper, the technique is applied to the
calculation of the excited states of conducting polymers. One
advantage of the technique is that it works on {\bf r} instead
of {\bf k} space.
This enables one to generate with increasing accuracy the local
correlation hole around an electron in the conduction band and
a hole in the valence band and to interpret physically the
different contributions to it. They are carefully studied both
for an on-site Hubbard type interaction and for long-range
interactions. The polarizability of $\sigma$
electrons enters the present calculations as an additional
screening mechanism of the $\pi$ electrons.

This paper is organized as follows. In the next section, we describe
the model Hamiltonian which we use in this work. In Sec.III, the
projection technique for excited states is described and the
excitation energies are given. The competition between a {\it loss}
of the correlation energy in the ground state and a {\it gain} of
the polarization energy is shown. In Sec.IV, we give the correlation
gap of conducting polymers under the scheme of the projection
technique. In Sec.V, the numerical results are given, both for the
Hubbard and long-range interactions. They are compared with the
exact solution at the undimerized and the independent dimer limits.
In the last section, we discuss the results we obtained and  the
screening mechanism of $\pi$ electron interactions for
{\it trans}-polyacetylene. In the appendixes, we give some details
of evaluating and the gap corrections by second-order perturbation
theory.

\section{MODEL HAMILTONIAN}
The $\pi$-electron system of one-dimensional conducting polymers
can be described by the Pariser-Parr-Pople (PPP) model \cite{oks},
\begin{equation}
H=H_{\rm SSH}+H_{\rm inter},
\end{equation}
where,
\begin{equation}
H_{\rm SSH}=-\sum_{l,s}[t_0-\alpha(u_{l+1}-u_l)](c_{l+1s}^\dagger
c_{ls}+h.c.)+\case{1}/{2}K\sum_l(u_{l+1}-u_l)^2
\end{equation}
is the Su-Schrieffer-Heeger (SSH) Hamiltonian \cite{ssh},
$c^\dagger_{ls}$ and $c_{ls}$
are electron creation and annihilation operators at site $l$ and with
spin $s$($=\alpha,\beta$), $u_l$ is the deviation of the $l$-th site
from the equilibrium position (with equal distance between sites),
the nearest-neighbor hopping integral between $l$- and
$l+1$-th sites has been taken as a linear function of the bond
modulation $u_{l+1}-u_l$, and the last term in Eq.(2.2) is the
elastic energy of this one-dimensional lattice. Furthermore,
\begin{equation}
H_{\rm inter}=\case{1}/{2}U\sum_l\rho_{l}\rho_{l}+\case{1}/{2}
\sum_{l,l'}{'}V_{ll'}\rho_{l}\rho_{l'}
\end{equation}
is the electron-electron interaction, which includes an on-site
Hubbard term $U$ and off-site interactions $V_{ll'}$ between
electrons on site $l$ and $l'$, and $\rho_{l}=c^\dagger_{l\alpha}
c_{l\alpha}+c^\dagger_{l\beta}c_{l\beta}-1$ the net charge density.
The prime in the second summation term implies $l\neq l'$.

For the calculation of ground state, an effective short-range
potential, {\it i.e.} Hubbard or extended Hubbard model (which
includes nearest-neighbor repulsion) is commonly used
to account for $\pi$-electron interaction in polymers. However,
when the system is away from the fully-shell electron distribution,
long-range polarizations become important. So it is better to use
a long-range potential for a correct description of $\pi$-electron
interactions in excited states. The interaction range depends on
the screening by the polarizability of
$\sigma$ electrons in the system. Various empirical
formulas \cite{oks} for this long-range potential of $\pi$ electrons
have been worked out. Here we use Ohno's formula \cite{ohno,oks}
for the interaction $V_{ll'}$,
which is given by
\begin{equation}
V_{ll'}={U\over{\sqrt{1+(r_{ll'}/r_0)^2}}}.
\end{equation}
Here, $r_0\approx 1.29$\AA, $r_{ll'}$ is the distance between
sites $l$ and $l'$ in unit of \AA\ and $U=11.13$ eV.
Roughly speaking, for polyacetylene, the carbon bond angle is
120$^\circ$ and average distance between the nearest-neighbor sites
is 1.4\AA.

For a dimerized lattice ($u_l=\pm u$), the SSH Hamiltonian can
be diagonalized exactly. Within Hartree-Fock approximation, we
can obtain a self-consistent solution by including the
exchange parts of the interaction (2.3). However, instead of a
self-consistent HF Hamiltonian, we simply start from
the following effective single-particle Hamiltonian,
\begin{equation}
H_0=-t\sum_{ls}[1\pm z](c^\dagger_{l+1s}c_{ls}+h.c.),
\end{equation}
where, $4t$ is the energy band-width and $z$ is the gap
order parameter, $+$ for long bonds and $-$ for short bonds.
Physically, there are two factors to influence the values
of these two effective quantities $t$
and $z$, besides the bare single-particle contribution,
which gives $t_0$ and $2\alpha u/t_0$ respectively. One
is the exchange-interaction contribution, which results from
off-site interactions. For the extended Hubbard model,
this exchange-interaction gives
\begin{mathletters}
\begin{eqnarray}
t&=&t_0+V\overline{m},\\
z&=&{{2\alpha u+V\delta m}\over{t_0+V\overline{m}}},
\end{eqnarray}
\end{mathletters}
where $\overline{m}$ and $\delta m$ are defined by $\langle
c^\dagger_{ls}c_{l+1s}\rangle=\overline{m}\pm \delta m$ \cite{hhkn}.
For a long-range interaction, we only consider an effective
change on the nearest-neighbor hopping instead of explicitly
including these hopping terms beyond the nearest-neighbors \cite{ph}.
Other is the correlation contribution \cite{dbkm}, which reduces the
band-width and increases the gap order parameter. While the treatment
of exchange contributions is Hartree-Fock approximation, the inclusion
of correlation contributions implies that we have gone beyond the mean
field theory. Mathematically, $t$ and $z$ are determined by
minimization of the total (both mean-field and correlation) energy
expectation value of the Hamiltonian (2.1) in the ground state.
It has been known \cite{dbkm} that the correlation contribution to the
parameter $z$ becomes important and makes the total energy much lower
in the interaction region ($U>2t$), and with this contribution the
calculation using local ansatz is valid until intermediate interaction
($U\sim 4t$).

In the Hamiltonian (2.5) we divide the operator $c_{ls}$
into two parts, {\it i.e.}, the electron operator in conduction
band $a_{ls}$ and the hole operator in valence band $b_{ls}$,
\begin{equation}
c_{ls}=a_{ls}+b^\dagger_{l\bar{s}},
\end{equation}
where,
\begin{mathletters}
\begin{eqnarray}
a_{ls}&=&{1\over{\sqrt{N}}}\sum_ke^{ikl}[i\sin\phi_k
+(-1)^l\cos\phi_k]a_{ks},\\
b^\dagger_{l\bar{s}}&=&{1\over{\sqrt{N}}}\sum_ke^{ikl}
[\cos\phi_k+i(-1)^l\sin\phi_k]b^\dagger_{-k\bar{s}},
\end{eqnarray}
\end{mathletters}
$\bar{s}=-s$, the summation is over a reduced Brillouin
zone $(-\pi/2,\pi/2]$, and the lattice constant has been taken
as unit. The Hamiltonian (2.5) becomes
\begin{equation}
H_0=2t\sum_{k,s}\epsilon(k)(a^\dagger_{ks}a_{ks}
+b^\dagger_{ks}b_{ks}-1),
\end{equation}
and the electron energy spectrum (in unit of $2t$)
\begin{equation}
\epsilon(k)=\sqrt{\cos^2k+z^2\sin^2k}.
\end{equation}
The transformation angle $\phi_k$ is defined by
\begin{equation}
\tan\phi_k={{z\sin k}\over{\epsilon(k)+\cos k}}.
\end{equation}
For a half-filling case, the ground state $|0\rangle$
is the vacuum of the electron operator $a_{ks}$ and
hole operator $b_{ks}$, so that it is also the vacuum
of operators $a_{ls}$ and $b_{ls}$.

We define the correlation functions $P_{ll'}$ and
$Q_{ll'}$ \cite{ph} as follows
\begin{mathletters}
\begin{eqnarray}
P_{ll'}&=&\langle c^\dagger_{ls}c_{l's}\rangle=\langle b_{l\bar{s}}
b^\dagger_{l'\bar{s}}\rangle,\\
Q_{ll'}&=&\langle c_{ls}c^\dagger_{l's}\rangle=\langle a_{ls}a^\dagger
_{ls}\rangle,
\end{eqnarray}
\end{mathletters}
and the anticommutators for the operators $a$ and $b$ are
\begin{mathletters}
\begin{eqnarray}
\{a_{ls},a^\dagger_{l's'}\}&=&\delta_{ss'}Q_{ll'},\\
\{b_{ls},b^\dagger_{l's'}\}&=&\delta_{ss'}P_{ll'}.
\end{eqnarray}
\end{mathletters}
Explicit expressions for these functions are given in Appendix A.
We will need them in the calculations of the following sections.

Substituting Eq.(2.7) into the interaction (2.3), we have the
two-particle part as follow \cite{hhkn},
\begin{equation}
H_1=\sum_mV_mO_m,
\end{equation}
where,
\begin{equation}
O_m=O_m^{(1)}+(O_m^{(2)}+h.c)+(O_m^{(3)}+h.c),
\end{equation}
and
\begin{mathletters}
\begin{eqnarray}
O_m^{(1)}&=\sum_{lss'}\{&(a^\dagger_{ls}a^\dagger_{l+ms'}
a_{l+ms'}a_{ls}+b^\dagger_{ls}b^\dagger_{l+ms'}
b_{l+ms'}b_{ls})\nonumber\\
&&-(a^\dagger_{ls}b^\dagger_{l+ms'}b_{l+ms'}a_{ls}
+a^\dagger_{l+ms}b^\dagger_{ls'}b_{ls'}a_{l+ms})\nonumber\\
&&+(a^\dagger_{ls}b^\dagger_{l\bar{s}}b_{l+m\bar{s'}}a_{l+ms'}
+a^\dagger_{l+ms}b^\dagger_{l+m\bar{s}}b_{l\bar{s'}}a_{ls'})\}
\end{eqnarray}
is the interaction between electron-electron, hole-hole and electron-hole;
\begin{equation}
O_m^{(2)}=\sum_{lss'}a^\dagger_{ls}a^\dagger_{l+ms'}
b^\dagger_{l+m\bar{s'}}b^\dagger_{l\bar{s}}
\end{equation}
is the spontaneous creation of two electron-hole pairs and
\begin{eqnarray}
O_m^{(3)}&=\sum_{lss'}&(a^\dagger_{ls}a^\dagger_{l+ms'}
b^\dagger_{l+m\bar{s'}}a_{ls}
+a^\dagger_{l+ms}a^\dagger_{ls'}b^\dagger_{l\bar{s'}}
a_{l+ms}\nonumber\\
&&+b^\dagger_{ls}b^\dagger_{l+ms'}a^\dagger_{l+m\bar{s'}}b_{ls}+
b^\dagger_{l+ms}b^\dagger_{ls'}a^\dagger_{l\bar{s'}}b_{l+ms})
\end{eqnarray}
\end{mathletters}
is the creation of an electron-hole pair through the scattering of an
electron or a hole. We have taken the translation invariance for
the interaction coefficients, {\it i.e.}, $V_m=V_{l,l+m}$ for $m\neq
0$ and $V_0=U/2$. The one-particle part of interaction (2.3)
has been absorbed into the effective Hamiltonian (2.5) as the exchange
contribution.

\section{PROJECTION TECHNIQUE}
We consider the case an electron or hole is added to the
half-filled system. The excitation energy is defined by
\begin{equation}
\varepsilon(k)=E_1(k)-E_0
\end{equation}
and is contained in the one-particle correlation function
\begin{equation}
R_s(k,\tau)=\langle g|c_{ks}e^{-\tau(H-E_0)}c^\dagger_{ks}|g\rangle.
\end{equation}
Here, $|g\rangle$ is the ground state of the half-filled system,
$E_0$ is the corresponding energy, $c^\dagger_{ks}$ creates an
electron ($a^\dagger_{ks}$) or a hole ($b^\dagger_{ks}$), and $E_1(k)$
is the energy for the system with the additional particle. Within
the quasiparticle approximation, the correlation function reduces to
\[R_s(k,\tau)=\langle g|c_{ks}c^\dagger_{ks}|g\rangle
e^{-\tau \varepsilon (k)}.\]
The quasiparticle energy $\varepsilon (k)$ shows up as
a pole of its Laplace transformation $R_s(k,x)$, which has been
shown \cite{bb} to be
\begin{equation}
R_s(k,x)=(c^\dagger_{ks}\Omega|{1\over{x-(L_0+H_1)}}
c^\dagger_{ks}\Omega).
\end{equation}
In the above, we have used a bilinear form in Liouville space
\begin{equation}
(A|B)=\langle 0|A^\dagger B|0\rangle^c,
\end{equation}
the index $c$ expresses the cumulant production \cite{pf}, which is
defined in Appendix B.
The Liouville operator $L_0$ acting on any operator $A$ gives
\begin{equation}
L_0A=[H_0,A]_-,
\end{equation}
and
\begin{equation}
\Omega=\lim_{x\rightarrow 0}(1+{1\over{x-(H_1+L_0)}}H_1),
\end{equation}
which transforms the ground state of $H_0$ ($|0\rangle$)
into the exact ground state ($|g\rangle$) \cite{pf}.

Now, we divide the Liouville space $R$ into a relevant part $R_0$ and
an irrelevant part $R_1$. Let $R_0$ be spanned by a set $\{A_\nu\}$ of
elements $|A_\nu)$. Then the projector \cite{pf}
\begin{equation} P=\sum_{\mu\nu}|A_\mu)\chi^{-1}_{\mu\nu}(A_\nu|
\end{equation}
projects onto the space $R_0$, and $\chi_{\mu\nu}=(A_\mu|A_\nu)$.

Within this relevant space $R_0$, we have
\begin{equation}
R_s(k,x)=\sum_{\mu,\nu}(c^\dagger_{ks}\Omega|A_\mu)\chi^{-1}_{\mu\nu}
\tilde{R}_\nu(k,x),
\end{equation}
where $\tilde{R}_\nu(k,x)$ denotes the set of functions
\begin{equation}
\tilde{R}_\nu(k,x)=(A_\nu|{1\over{x-P(L_0+H_1)}}Pc^\dagger_{ks}
\Omega).
\end{equation}
Then we have the equation
\begin{equation}
\sum_{\eta,\mu}(x\chi_{\nu\eta}-\omega_{\nu\eta})\chi^{-1}_{\eta\mu}
\tilde{R}_\mu(k,x)=(A_\nu|Pc^\dagger_{ks}\Omega)
\end{equation}
with
\begin{equation}
\omega_{\mu\nu}=(A_\mu|(L_0+H_1)A_\nu).
\end{equation}
The excitation energies $\varepsilon(k)$ are given by the poles of
$R_s(k,x)$ or $\tilde{R}_\nu(k,x)$, so are the solutions of the
following equation
\begin{equation}
 \det\{x\chi_{\mu\nu}-\omega_{\mu\nu}\}=0,
\end{equation}
if we choose the elements $A_\mu$, which span the relevant space
$R_0$, so that $(A_\mu|c^\dagger_{ks}\Omega)\neq 0$.

In the half-filled ground state, the electron distribution is
fully-shell. The correlation interaction (2.14) would give rise
to spontaneous excitations of electron-hole pairs. These spontaneous
excitations lower the total energy. In the presence of an additional
electron or hole, the correlation causes two processes: one is the
blocking of the spontaneous excitations of electron-hole pairs if
these excitations are associated with the extra particle; other is
the excitation of electron-hole pairs through the scattering of
the extra particle. The former (which we use operator
$S^{\eta}_\mu$ to describe) gives rise to a {\it loss} of the
correlation energy in ground state so that the excitation energy
is increased. And the latter (which we use $S^{\pi}_\mu$ to
describe) is a polarization process, which decreases the excitation
energy by the {\it gain} of the polarization energy.
These two processes induce a polarization cloud around
the extra particle so that an electron polaron or a hole
polaron is formed. Here we set $\{A_\mu\}=
\{c^\dagger_{ks}\}\oplus\{S^{\pi}_\mu
c^\dagger_{ks}\}\oplus\{c^\dagger_{ks}S^{\eta}_\mu\}$,
where $\{S^{\pi}_\mu\}$ are the operators, which do not
commute with the extra particle creation operator
$c^\dagger_{ks}$ while $\{S^{\eta}_\mu\}$ commute with
the operator $c^\dagger_{ks}$.
Then we can rewrite the excitation energy as two parts
\begin{equation}
\varepsilon(k)=2t\cdot\epsilon(k)+\varepsilon^{\rm corr}(k),
\end{equation}
{\it i.e.}, the mean-field energy spectrum and correlation
contribution. The latter is combined by two parts, too,
\begin{equation}
\varepsilon^{\rm corr}(k)=\varepsilon^{\pi}(k)
+\varepsilon^{\eta}(k),
\end{equation}
where
\begin{mathletters}
\begin{equation}
\varepsilon^{\pi}(k)=-\sum_\mu{\bf \pi}_\mu(k)
(c^\dagger_{ks}|H_1S^{\pi}_\mu c^\dagger_{ks})
\end{equation}
is the {\it gain} of the energy from the polarization
process with the extra particle, and
\begin{equation}
\varepsilon^{\eta}(k)=\sum_\mu{\bf
\eta}_\mu(k)(c^\dagger_{ks}|H_1c^\dagger_{ks}S^{\eta}_\mu)
\end{equation}
\end{mathletters}
is the {\it loss} of the correlation energy in the ground
state due to the extra particle.

When the mutual influence of the two types correlation is
neglected by set $(S^\pi_\mu c^\dagger_{ks}|H_1c^\dagger
_{ks}S^\eta_\nu)\simeq 0$, which we believe is small and is
negligible for a non-strong interaction (see Appendix C),
then we can write down
\begin{mathletters}
\begin{eqnarray}
{\bf \pi}_\mu(k)&=&\sum_\nu C^{-1}_{\mu\nu}(\pi)
(S^\pi_\nu c^\dagger_{ks}|H_1c^\dagger_{ks}),\\
{\bf \eta}_\mu(k)&=&\sum_\nu C^{-1}_{\mu\nu}(\eta)
(c^\dagger_{ks}S^\eta_\nu|H_1c^\dagger_{ks}).
\end{eqnarray}
\end{mathletters}
The coefficients are the inversions of matrices $C(\pi)$ and
$C(\eta)$, which are given by
\begin{mathletters}
\begin{eqnarray}
C_{\mu\nu}(\pi)&=&(S^\pi_\mu c^\dagger_{ks}|[L_0+H_1
-\varepsilon(k)]S^\pi_\nu c^\dagger_{ks}),\\
C_{\mu\nu}(\eta)&=&-(c^\dagger_{ks}S^\eta_\mu|[L_0+H_1
-\varepsilon(k)]c^\dagger_{ks}S^\eta_\nu).
\end{eqnarray}
\end{mathletters}
When the mutual influences of the correlated excitations
and the polarized scatterings are taken into account, the
expressions for ${\bf \pi}_\mu(k)$ and ${\bf \eta}_\mu(k)$
become somewhat more complex.

{}From the expressions (3.17), we notice that both the {\it loss}
of the correlation energies and the {\it gain}
of polarization energies are dependent \cite{pf} of the exact
excitation energy $\varepsilon(k)$. This dependence comes from the
non-zero of susceptibilities $\chi_{\mu\nu}$.
It makes the excited energies must be obtained self-consistently and
results in a narrowing of energy bands \cite{baf}. The more
important is that, we will see in Sec.V, the dependence ensures
that its results are correct for intermediate interaction
strength.

\section{CORRELATION GAP}
A band gap in a semiconductor is the energy it costs to move an
electron on the top of valence band to the bottom of conduction
band. In an electron-hole picture it is the minimum energy to create
an electron-hole pair. Here, we define the {\it correlation gap}
$E_{\rm CG}$ as a contribution of the correlation
effect, {\it i.e.}, it is the difference between the exact band gap and
the mean-field band gap
\begin{equation}
E_{\rm CG}=E_{\rm gap}-E_{\rm gap}^{\rm HF}.
\end{equation}
For our case where there exists the symmetry of electrons
and holes, the excitation energies of an electron polaron and
a hole polaron are same. So the {\it correlation gap} should
be twice of the excitation energies of the polaron
$\varepsilon^{\rm corr}(k_0)$ ($k_0=\pi/2$), if the interaction
between the electron polaron and hole polaron ---
the excitonic effect --- is neglected. We leave this excitonic
effect in this system as a separate study.

For the calculation of this excitation energy, we should first
choose the polarized scattering operator $S^\pi_\mu$ and the
correlated excitation operator $S^\eta_\mu$. When the
interaction is not strong, we can choose
\begin{mathletters}
\begin{equation}
S^\pi_{ij}=\case{1}/{2}\sum_{\sigma,\sigma'}a^\dagger_{i\sigma}
b^\dagger_{i\bar{\sigma}}a^\dagger_{j\sigma'}a_{j\sigma'}
\end{equation}
and
\begin{equation}
S^\eta_{ij}=\case{1}/{2}\sum_{\sigma,\sigma'}a^\dagger_{i\sigma}
b^\dagger_{i\bar{\sigma}}a^\dagger_{j\sigma'}b^\dagger_{j\bar{\sigma'}}.
\end{equation}
\end{mathletters}
These operators span a relevant space $R_0$ for the case an electron
is added into the conduction band. The physical meaning of the two
kinds of operators has been discussed in the previous section.

The choice of the above operators means that we do not break the
spin symmetry. It is correct for a dimerized system, where
mean-field ground state is of bond order waves (BOW) or charge
density wave (CDW), without spin density waves (SDW). However,
for a uniform lattice --- the undimerized limit --- the
SDW would play a non-negligible role. It leads the calculation
using the above operators underestimate the energy gap. Furthermore,
we have only considered two electron-hole pairs excitations, which
is enough only for a non-strong interaction. As shown in the calculation
for ground state, it works until $U\sim 4t$ \cite{dbkm}. This is
enough for the polymers like polyacetylene, where $U\sim 4t_0$.
We will see it also holds for excited states in the following
section. However, for $U>4t$, the system will undergo a
spin-Peierls transition, there the spin-spin interaction
is dominant. This is not the regime we interest in this paper.

With the translation invariance, we reduce the dimension of
relevance space $R_0$ by using the following operators,
\begin{mathletters}
\begin{eqnarray}
S^\pi_d&=&\sum_iS^\pi_{i,i+d},\\
S^\eta_d&=&\sum_iS^\eta_{i,i+d}.
\end{eqnarray}
\end{mathletters}
Now we write down the following four kinds of matrices:
1. susceptibility-matrix
$(S^\pi_da^\dagger_{ks}|S^\pi_{d'}a^\dagger_{ks})$,
$(S^\eta_da^\dagger_{ks}|S^\eta_{d'}a^\dagger_{ks})$;
2. projection matrix
$(a^\dagger_{ks}|H_1S^\pi_da^\dagger_{ks})$,
$(a^\dagger_{ks}|H_1S^\eta_da^\dagger_{ks})$;
3. hopping matrix
$(S^\pi_da^\dagger_{ks}|H_0S^\pi_{d'}a^\dagger_{ks})$,
$(S^\eta_da^\dagger_{ks}|H_0S^\eta_{d'}a^\dagger_{ks})$
and 4. interaction matrix
$(S^\pi_da^\dagger_{ks}|H_1S^\pi_{d'}a^\dagger_{ks})$,
$(S^\eta_da^\dagger_{ks}|H_1S^\eta_{d'}a^\dagger_{ks})$.
They should be evolved before the calculation of
excitation energies. These matrices are given in Appendix B.

\section{NUMERICAL RESULTS}
\subsection{Hubbard interaction}

We consider Hubbard interaction in this subsection based
on the general formula established in the previous two sections.
For such a short-range interaction, we first choose only on-site
local operators $S^\pi_d$ and $S^\eta_d$ with $d=0$ as the
elements of set $\{A_\mu\}$. With this choice, we have the
expressions for excitation energies
\begin{mathletters}
\begin{eqnarray}
\varepsilon^\pi(k_0)&=&-{a_\pi\over{b_\pi-\varepsilon(k_0)}}
\cdot U^2,\\
\varepsilon^\eta(k_0)&=&{a_\eta\over{b_\eta-\varepsilon(k_0)}}
\cdot U^2,
\end{eqnarray}
\end{mathletters}
where, the coefficients $a$ and $b$ are as follows
\begin{mathletters}
\begin{eqnarray}
a_\pi&=&\sum_lR_lP_l^3,\\
b_\pi&=&3\sum_lE_lR_lP^2_l/a_\pi;
\end{eqnarray}
\end{mathletters}
and
\begin{mathletters}
\begin{eqnarray}
a_\eta&=&\sum_l(-1)^lR_lP_l^3,\\
b_\eta&=&4t\epsilon(k_0)+3\sum_l(-1)^lE_lR_lP_l^2/a_\eta.
\end{eqnarray}
\end{mathletters}
The functions $P_l$, $E_l$ and $R_l$ are defined in
Appendix A. The coefficients $b_\pi$ and $b_\eta$ are independent
of Hubbard strength $U$ because the interaction matrix vanishes
here.

The equation (5.1) combined with the equations (3.14) and (3.15)
determines the excitation energy $\varepsilon^{\rm corr}(k_0)$ as
well as the correlation gap $E_{\rm CG}$ self-consistently.
For {\it trans}-polyacetylene, the band gap has been measured as
1.8 eV \cite{nsea}. For Hubbard interaction, we will see the
correlation gap $E_{\rm CG}$ is always positive, {\it i.e.}, the
band gap is enhanced by the Hubbard correlation.
At $z=0.15$, which opens a band gap of 1.5 eV in the single-particle
energy spectrum if $t=2.5$ eV as taken by SSH \cite{ssh}, we have
$a_\pi=0.055, a_\eta=0.196, b_\pi=4.34t$,
and $b_\eta=4.43t$. The correlation gap $E_{\rm CG}$ obtained
is plotted as a function of $U$ in Fig.1 as a dashed line, while
the result obtained with off-site operators $d$ up to $10$ is
shown as a solid line. As to make a comparison, we have shown the
second-order perturbation result as a dotted line and the Monte
Carlo results \cite{jh} as points with error bars in Fig.1.

First let's see the limit of small $U$, where the correlation gap
is of the form $\kappa U^2/t$. The result with on-site operators
gives $\kappa=0.0677$, which is about $90\%$ of the value
$\kappa=0.0755$ calculated from the second-order perturbation theory
(see Appendix C). Here we should mention that the second-order
perturbation theory for the correlation gap is exact in the limit
of small $U$ for the case a finite dimerization has opened a gap.
It is different with the uniform lattice ($z=0$) case where the
band gap is purely Coulomb gap \cite{lw} and the perturbation
theory doesn't work. Our result can be improved greatly by include
of off-site operators, for $d$ up to $10$, we obtain
$\kappa=0.0723$, which reaches $96\%$. From Fig.1, we can see
that the off-site operators become more important when the
interaction strength $U$ goes larger, which is easy to be
understood.

At $z=0$, the undimerized limit, the exact gap is
available from the paper of Lieb and Wu \cite{lw},
\begin{equation}
E_{\rm CG}=U-4t+8\sum_{n=1}^\infty
(-1)^n[\case{1}/{2}nU-(t^2+\case{1}/{4}n^2U^2)^{1/2}],
\end{equation}
which is shown in Fig.1 as a dot-dashed line. Compared with
this exact solution, the correlation gap at $z=0.15$ obtained by
our projection technique is always larger even only with
the on-site operators. It means the correlation gap is
enhanced by a finite dimerization. However, the perturbation
theory shows us that there is a critical interaction strength
$U_c$, when $U$ is larger than $U_c$, the dimerization reduces
the correlation gap while it enhances the correlation gap when
$U$ is smaller than $U_c$. Obviously this is an artifact. The
fact tells us it is important that the dependence of both the
correlation energy and the polarization energy on the exact
excited energy \cite{hhf} because it is the dependence which raises
the correlation gap.

In Fig.2, we show the dependence of the correlation gap on the
dimerization $z$ for different interaction strengths $U$. At
very small dimerization $z$, as declared previously, the result
we obtained here underestimates the correlation gap because the choice
of the operator set (4.3), which doesn't break the spin symmetry.
Comparing with the exact solution at $z=0$, we can see that this
region is very small for small $U$, and it is about $0.1$
for $U=4t$. The enhancement of the correlation gap by a small
dimerization is shown clearly and the correlation gap reaches its
maximum around $z=0.3$. For a larger $z$, the correlation gap decreases
monotonously. This behavior is easy to be understood by compared
with the correlation energy in the ground state \cite{ph}, which is decreased
by a small dimerization and after reach its minimum value it goes
up monotonously with the increase of the dimerization. The reason
is that for Hubbard interaction, the {\it loss} of correlation energy
is dominant and the polarization energy is small.

At $z=1$, the independent dimer limit, the exact
correlation gap can be obtained easily,
\begin{equation}
E_{\rm CG}=(U^2+64t^2)^{1/2}-8t,
\end{equation}
which has been indicated in Fig.2 as well as the exact
solution (5.4) at $z=0$. At the independent dimer limit, the
long-range operators are not important while the high-order
excitations become important for a strong interaction. Since
the calculation only with low-order excitations gives a value
larger than the exact result, the high-order excitations should
reduce the correlation gap, which is contrary to the effect of
long-range operators. So we expect that there are some
cancellations between the longer-range excitations and
the high-order excitations.

In the end of this subsection, we mention the work of Sun
{\it et al.} \cite{xsea}, which used a screened Coulomb interaction and
found the optical gap is reduced by the electron interaction if
the screening is very strong. This result is not contrary to ours
because as stated in their paper this reduction is due to the
non-diagonal bond-charge repulsion, which we don't include in
this paper.

\subsection{Long-range interaction}

For the long-range interaction, we take the Ohno formula (2.4) to
parameterize the interaction coefficient $V_{ll'}$, which is
dependent of the distance between the site $l$ and $l'$. With
a dimerization $u$, the change of bond lengths is $4u/\sqrt{3}$
and the bond alternation is formed. Within the mean-filed
approximation and only nearest-neighbor-site interaction, the
relation of the dimerization $u$ and the gap parameter $z$ is
shown in Eq.(2.6). With the longer-range interaction as well as the
effect of correlation, the dependence of $u$ and $z$ should
be determined numerically as discussed in Sec.II.

Due to the existence of the long-range interaction, the relevant
space $R_0$ only with the on-site operators ($d=0$) is too small to give
a qualitative correct result, the inclusion of long-range operators is
necessary because the polarization process is of a much longer
correlation length than that of correlated excitations. With the
relevant space $R_0$ spanned by the operators (4.3) with $d$ up to 10,
the correlation gap is shown in Fig.3(a). We can see that it is
the long-range interaction which makes the {\it gain} of the polarization
energy becomes dominant so that the correlation gap is negative.
This result means the long-range correlation reduces the mean-field
energy gap, which is unphysically large from the {\it ab initio}
self-consistent filed (SCF) calculation \cite{gkgs}. This reduction
becomes larger with the increase of a small dimerization. At about
$z=0.1$, the correlation gap has reached its minimum and then increases
with the increase of the dimerization.

At $z=0.2$, the reduction of the band gap is about 3.4 eV. For
{\it trans}-polyacetylene, the SCF gap is 6.9 eV \cite{gkgs}, among which
the SCF dimerization contributes 1.05 eV and the exchange
interaction contributes the rest. So the Coulomb gap arose
from the electron-electron interaction is 2.45 eV. The deviation
comes from the model interaction (2.3) which doesn't include the
polarizability of $\sigma$ electrons in the system. The presence
of $\sigma$ electrons would be favorable to the $\pi$ electron
excitations so that it should reduce the band gap further.

In a simple way, the polarizability of $\sigma$ electrons could be
considered as a screening of the $\pi$-electron interactions, {\it i.e.},
the Ohno interaction is screened as follows \cite{ph}
\begin{equation}
\tilde{V}_{ll'}={V_{ll'}\over{\epsilon_0}}.
\end{equation}
At this moment, the on-site Hubbard interaction $U$ is kept unchanged.
With the screen for $\epsilon_0>2$, the correlation gap becomes
positive again. It means that the screen suppresses the polarization
process so that the {\it gain} of polarization energy is reduced and
the {\it loss} of the correlation energy is again dominant. This is
same as the correlation gap for a short-range (Hubbard) interaction
qualitatively. Fig.3(b) shows the correlation gap for the screened
interaction with $\epsilon_0=3$.

The screening reduces the $\pi$ electron interaction range so that
the correlation increases the gap. But at the same time, the
screening reduces the exchange contribution, too, by a factor of
$1/\epsilon_0$. In fact the screen decreases the energy gap,
{\it i.e.}, the contribution of the $\sigma$ electrons tends
to reduce the band gap.

\section{DISCUSSIONS AND SUMMARY}

First let's see the {\it ab initio} work of K\"onig and Stollhoff
\cite{gkgs}. At SCF level, it gives the SCF energy gap 6.9 eV
at the equilibrium dimerization $u=3.28$ pm. The equilibrium
dimerization $u$ decreases to 2.52 pm when electron correlations
are included.  By fitting the ground state energy, they obtained
the model parameters $\alpha=40$ meV/pm, $K=3.9$ meV/pm$^2$
for $t_0=2.5$ eV. With the above parameters, we obtain the SCF
gap of 7.0 eV for the long-range interaction (2.4). Among it,
the dimerization contributes $8\alpha u=1.05$ eV while the
exchange contribution $2\sum_l(-1)^{-1}V_{2l+1}P_{2l+1}$ is
5.95 eV. The agreement with {\it ab initio} calculation shows that
the PPP model (2.1) describes the $\pi$-electrons in conducting
polymers quite well.

With electron correlations, the bond alternation is reduced
\cite{bhwsn}. Then the dimerization contribution $8\alpha u$
to the band gap becomes 0.8 eV and the exchange interaction
contribution to the band gap decreases 5.83 eV. With the screen
(5.6), the exchange contribution to the gap is
2.49 eV for $\epsilon_0=2$ and is 1.52 eV for $\epsilon_0=3$.
For the screen with a fixed on-site repulsion $U$
($=11.13$ eV), the correlation gap is 4.58 eV at
$\epsilon_0=\infty$ ({\it i.e.}, only Hubbard on-site
interaction and no exchange contribution), and $-3.39$ eV
at $\epsilon_0=1$ ({\it i.e.}, no screen). That gives
the band gap 2.9 eV at the limit of unscreening interactions
($\epsilon_0=1$) and 5.4 eV at the limit of completed screening
interactions ($\epsilon_0=\infty$). Obviously in this way,
the contribution of $\sigma$ electrons to the band gap
is contrary to the physical fact we expect, {\it i.e.},
the polarizability of $\sigma$ electrons should reduce the band gap.

We need to consider a more realistic situation of conducting
polymers: the on-site $\pi$ electron interaction is not completely
local, {\it i.e.}, it has a nonlocal part. So the on-site repulsion
$U$ would be screened by the polarizability of $\sigma$ electrons
from its ``bare'' value (11.13 eV).
We take the form $U=U_0+V$, where $U_0$ ($=3.6$ eV) is the
``net'' on-site repulsion and $V$ is the nearest-neighbor interaction,
which is screened by $\sigma$ electrons as in Eq.(5.6). In this way,
at the limit of $\epsilon_0=\infty$, the interaction
is the Hubbard model with $U_0$, the corresponding correlation
gap is 0.42 eV, and then the band gap is 1.22 eV. That shows
a reasonable reduction from the polarizability of $\sigma$ electrons.
To get the band gap 1.8 eV, we have to set $\epsilon_0=9$, a quite
strong screening to $\pi$-electrons, then the exchange
contribution is 0.43 eV and the correlation gap is 0.57 eV.
With the screen, the on-site repulsion $U=4.4$ eV and the
nearest-neighbor interaction $V=0.8$ eV, the interaction range
is very short.

The electron-electron interaction parameters $U$ and $V$
we determined from the energy gap are quite different
with the values obtained by fitting the groundstate energy curve
of the {\it ab initio} calculation on dimerization. In fact it
has appeared for many years that the discrepancy of the
strength of electron interactions among the theoretical results
based on different physical quantities. Some of the results are
tabulated in Table I. It seems that
it is not appropriate to talk about the strength of electron
interactions in polymers based on such a simple model.
The physical reason behind it is that the effective $\pi$ electron
interaction is strongly affected by the polarizability of $\sigma$
electrons,
which play a different role for different physical processes.
The interaction depends on physical processes. From Table I, we can
see that there are mainly two kinds of physical processes for
the interaction, one is in the ground state, other is in excited
states. On the ground state, the interaction is about 10 eV
\cite{dbkm,gkgs} while
it is about 4 eV \cite{skdh,wksk,eeea,jgea,ghs} on the excited states.

Finally, we summarize this work as follows. By using the projection
technique for excited states, we have studied the correlation
effects on the band gap of conducting polymers. In the presence
of an extra particle (electron or hole), the correlation
induces a polarization cloud around the extra particle, which
forms a polaron. For the excitation energy of a polaron,
there is a competition between a {\it loss} of the correlation
energy in the ground state and a {\it gain} of the polarization
energy. For a long-range interaction, the {\it gain} of
polarization energy is dominant and the correlation decreases
the band gap. A screening by the polarizability of $\sigma$
electrons reduces the interaction range and suppresses the
{\it gain} of the polarization energy due to the
{\it correlation}.  The {\it loss} of correlation
energy becomes dominant and the correlation increases the
band gap for the short-range or Hubbard interaction.
A small dimerization leads to a further enhancement (reduction)
of the correlation gap for the Hubbard (long-range) interaction.
For {\it trans}-polyacetylene, we obtain the on-site repulsion
$U=4.4$ eV and the nearest-neighbor interaction $V=0.8$ eV by
fitting the band gap to 1.8 eV. The screening of $\pi$ electrons
due to the polarizability of $\sigma$ electrons is quite strong.

\acknowledgements
The author would like to thank Professor P. Fulde for his
constant interest in this work and the critical reading of
the manuscript. He is grateful to K. Becker, P. Fulde,
P. Horsch, G. Stollhoff, X. Sun, and L. Yu for their
helpful discussions. He would also like to acknowledge
the financial support from the Max-Planck-Gesellschaft
and the hospitality of Professor P. Fulde.

\appendix{CORRELATION FUNCTIONS}
The correlation functions $P_{ll'}$ and $Q_{ll'}$ defined in
Eq.(2.12) have the relation
\begin{equation}
P_{ll'}+Q_{ll'}=\delta_{ll'}.
\end{equation}
For even value of $m$ and $n$,
\begin{mathletters}
\begin{eqnarray}
&&P_{m,n}=P_{m+1,n+1}=\case{1}/{2}\delta_{m,n},\\
&&P_{m+1,n}=P_{n,m+1}=P_{m-n+1},
\end{eqnarray}
\end{mathletters}
here
\begin{equation}
P_{2l+1}=\frac{1}{\pi}\int_0^{\pi/2}dk{{\cos k\cos (2l+1)k
+z\sin k\sin (2l+1)k}\over{\sqrt{\cos^2k+z^2\sin^2k}}}.
\end{equation}

The energy correlation function is defined as
\begin{eqnarray}
E_{ll'}&=&(c_{ls}|H_0c_{l's})=(b^\dagger_{l\bar{s}}|
H_0b^\dagger_{l'\bar{s}})\nonumber\\
&=&\langle b_{l\bar{s}}H_0b^\dagger_{l'\bar{s}}\rangle
-\langle H_0\rangle\langle b_{l\bar{s}}b^\dagger_{l'\bar{s}}
\rangle
\end{eqnarray}
and then
\begin{eqnarray}
\tilde{E}_{ll'}&=&(c^\dagger_{ls}|H_0c^\dagger_{l's})
=(a^\dagger_{ls}|H_0a^\dagger_{l's})\nonumber\\
&\equiv& (-1)^{l+l'}E_{ll'},
\end{eqnarray}
where, we have use the cumulant production. Similarly as did
for $P_{ll'}$, for even values of $m$ and $n$ we have
\begin{mathletters}
\begin{eqnarray}
&&E_{m,n}=E_{m+1,n+1}=E_{m-n},\\
&&E_{m+1,n}=E_{n,m+1}=E_{m-n+1},
\end{eqnarray}
\end{mathletters}
here,
\begin{equation}
E_{2l}=\case{1}/{2}t[(1+z)(P_{2l+1}+P_{-2l+1})+
(1-z)(P_{2l-1}+P_{-2l-1})],
\end{equation}
and
\begin{equation}
E_{2l+1}=\case{1}/{2}t[(1+z)\delta_{l,0}+(1-z)\delta_{l,-1}].
\end{equation}

The polarized correlation function $R_{ll'}(k)$ is defined by
\begin{equation}
(a^\dagger_{ks}|a^\dagger_{l's}a_{ls}a^\dagger_{ks})
=R_{ll'}(k)/N,
\end{equation}
$N$ is the number of sites. Similarly, we have
\begin{eqnarray}
&&(b^\dagger_{ks}|b^\dagger_{l's}b_{ls}b^\dagger_{ks})
=(-1)^{l+l'}R_{ll'}(k)/N,\\
&&R_{ll'}(k)=R_{l'l}^*(k).
\end{eqnarray}
For even numbers of $m$ and $n$,
\begin{mathletters}
\begin{eqnarray}
&&R_{m,n}(k)=R_{m+1,n+1}(k)=R_{m-n}(k),\\
&&R_{m+1,n}(k)=R_{m-n+1}(k),
\end{eqnarray}
\end{mathletters}
where,
\begin{mathletters}
\begin{eqnarray}
&&R_{2l}(k)=e^{2ikl},\\
&&R_{2l+1}(k)=-e^{ik(2l+1)-2i\phi_k}.
\end{eqnarray}
\end{mathletters}
At $k_0=\pi/2$, $R_{2l}=-R_{2l+1}=(-1)^l$, independent
of the dimerization $z$.

\appendix{COMPUTATIONS OF THE FOUR KINDS OF MATRICES}
The definition \cite{pf} of cumulant productions is as follow
\begin{equation}
\langle 0|A^\alpha_1\cdots A^\nu_n|0\rangle^c=
\frac{\partial^\alpha}{\partial \lambda^\alpha_1}\cdots
\frac{\partial^\nu}{\partial \lambda^\nu_n}\ln
\langle 0|\prod_{i=1}^ne^{\lambda_iA_i}|0\rangle|_{
\lambda_1=\cdots =\lambda_n=0}.
\end{equation}
Then we have the following properties:
\begin{mathletters}
\begin{eqnarray}
\langle 0|AB|0\rangle^c&=&\langle 0|AB|0\rangle-\langle 0|A|0\rangle
\langle 0|B|0\rangle,\\
\langle 0|ABC|0\rangle^c&=&\langle 0|ABC|0\rangle-\langle 0|A|0\rangle
\langle 0|BC|0\rangle^c-\langle 0|B|0\rangle\langle
0|AC|0\rangle^c\nonumber\\
&&-\langle 0|C|0\rangle\langle 0|AB|0\rangle^c-\langle 0|A|0\rangle
\langle 0|B|0\rangle\langle 0|C|0\rangle,\\
&& ...... \nonumber
\end{eqnarray}
\end{mathletters}

In terms of the properties of cumulants and the correlation
functions in Appendix A, we compute the four kinds
of matrices as follows.\\
1. Susceptibility matrix:
\begin{mathletters}
\begin{eqnarray}
(S^\pi_{ij}a^\dagger_{ks}|S^\pi_{i'j'}a^\dagger_{ks})&=&
-\case{1}/{4N}R_{ii'}P_{jj'}(2Q_{ii'}Q_{jj'}-Q_{ij'}Q_{ji'}),\\
(S^\eta_{ij}a^\dagger_{ks}|S^\eta_{i'j'}a^\dagger_{ks})&=&
\case{1}/{4N}Q_{ii'}R_{jj'}(2P_{ii'}P_{jj'}-P_{ij'}P_{ji'})\nonumber\\
&&+(i\rightleftharpoons j)+(i'\rightleftharpoons j')
+(i,i'\rightleftharpoons j,j').
\end{eqnarray}
\end{mathletters}
2. Projection matrix:
\begin{mathletters}
\begin{eqnarray}
(a^\dagger_{ks}|O_mS^\pi_{ij}a^\dagger_{ks})&=&
2\sum_l(S^\pi_{ll+m}a^\dagger_{ks}|S^\pi_{ij}a^\dagger_{ks})
+(m \rightleftharpoons -m),\\
(a^\dagger_{ks}|O_mS^\eta_{ij}a^\dagger_{ks})&=&
2\sum_l(S^\eta_{ll+m}a^\dagger_{ks}|S^\eta_{ij}a^\dagger_{ks}).
\end{eqnarray}
\end{mathletters}
3. Hopping matrix:
\begin{mathletters}
\begin{eqnarray}
(S^\pi_{ij}a^\dagger_{ks}|H_0S^\pi_{i'j'}a^\dagger_{ks})&=&
-\case{1}/{4N}\{R_{ii'}P_{jj'}(2\tilde{E}_{ii'}Q_{jj'}
+2Q_{ii'}\tilde{E}_{jj'}-Q_{ij'}\tilde{E}_{ji'}
-\tilde{E}_{ij'}Q_{ji'})\nonumber\\
&&\ \ \ \ \ \ +R_{ii'}E_{jj'}(2Q_{ii'}Q_{jj'}-Q_{ij'}Q_{ji'})\},\\
(S^\eta_{ij}a^\dagger_{ks}|H_0S^\eta_{i'j'}a^\dagger_{ks})&=&
4t\epsilon_k(S^\eta_{ij}a^\dagger_{ks}|S^\eta_{i'j'}a^\dagger_{ks})\nonumber\\
&&+\{\case{1}/{4N}[(Q_{ii'}R_{jj'}+R_{ii'}Q_{jj'})
(2E_{ii'}P_{jj'}-E_{ij'}P_{ji'})\nonumber\\
&&\ \ \ \ \ \ \ \ +R_{ii'}\tilde{E}_{jj'}(4P_{ii'}P_{jj'}
-2P_{ij'}P_{ji'})]\nonumber\\
&&\ \ +(i\rightleftharpoons j)+(i'\rightleftharpoons j')
+(i,i'\rightleftharpoons j,j')\}.
\end{eqnarray}
\end{mathletters}
4. Interaction matrix:
\begin{mathletters}
\begin{eqnarray}
(S^\pi_{ij}a^\dagger_{ks}|O_mS^\pi_{i'j'}a^\dagger_{ks})&
=-\case{1}/{8N}R_{ii'}\sum_l&\{
P_{jj'}Q_{li'}Q_{l+mj'}(2Q_{li}Q_{l+mj}-Q_{lj}Q_{l+mi})\nonumber\\
&&+P_{jj'}Q_{lj'}Q_{l+mi'}(2Q_{lj}Q_{l+mi}-Q_{li}Q_{l+mj})\nonumber\\
&&-2P_{l+mj}P_{l+mj'}Q_{li'}(2Q_{li}Q_{jj'}-Q_{lj}Q_{ij'})\nonumber\\
&&-2P_{l+mj}P_{l+mj'}Q_{lj'}(2Q_{lj}Q_{ii'}-Q_{li}Q_{ji'})\nonumber\\
&&+4P_{lj}P_{l+mj'}Q_{l+mj'}(2Q_{lj}Q_{ii'}-Q_{li}Q_{ji'})\nonumber\\
&&-2P_{lj}P_{l+mj'}Q_{l+mi'}(2Q_{lj}Q_{ij'}-Q_{li}Q_{jj'})\nonumber\\
&&+(m \rightleftharpoons -m)\},
\end{eqnarray}
\begin{eqnarray}
(S^\eta_{ij}a^\dagger_{ks}|O_mS^\eta_{i'j'}a^\dagger_{ks})&
=\case{1}/{4N}\sum_l&\{(R_{li}Q_{li'}+Q_{li}R_{li'})
Q_{l+mj}Q_{l+mj'}(2P_{ii'}P_{jj'}-P_{ij'}P_{ji'})\nonumber\\
&&+Q_{ii'}R_{jj'}P_{li'}P_{l+mj'}(2P_{li}P_{l+mj}-P_{lj}P_{l+mi})
\nonumber\\
&&-[(R_{l+mi}Q_{l+mi'}+Q_{l+mi}R_{l+mi'})Q_{jj'}+Q_{l+mi}Q_{l+mi'}
R_{jj'}]\times\nonumber\\
&&\ \ \times [P_{li'}(2P_{li}P_{jj'}-P_{lj}P_{ij'})
+P_{lj'}(2P_{lj}P_{ii'}-P_{li}P_{ji'})]\nonumber\\
&&+[(R_{li}Q_{l+mi'}+Q_{li}R_{l+mi'})Q_{jj'}+Q_{li}Q_{l+mi'}
R_{jj'}]\times\nonumber\\
&&\ \ \times [2P_{l+mi'}(2P_{li}P_{jj'}-P_{lj}P_{ij'})
-P_{l+mj'}(2P_{li}P_{ji'}-P_{lj}P_{ii'})]\nonumber\\
&&+(m\rightleftharpoons -m)\}\nonumber\\
&+(i\rightleftharpoons&j)
+(i'\rightleftharpoons j')+(i,i'\rightleftharpoons j,j').
\end{eqnarray}
\end{mathletters}
The matrices with the translational invariance are easily got
from the above matrices.

\appendix{SECOND-ORDER PERTURBATION THEORY}
By use of the second-order perturbation theory, the correlation energy
for the state with an extra electron in the half-filled ground state
is
\begin{equation}
E_1^{\rm corr}(k)=-\sum_m{|\langle m|H_1a^\dagger_{ks}|0\rangle|^2
\over{E_m-E_k}}.
\end{equation}
With the equation (2.14), we have the Hubbard interaction as
\begin{equation}
H_1=\case{1}/{2}U\{O_0^{(1)}+O_0^{(2)}+O_0^{(2)\dagger}
+O_0^{(3)}+O_0^{(3)\dagger}\},
\end{equation}
and
\begin{equation}
H_1a^\dagger_{ks}|0\rangle=\case{1}/{2}U\{O_0^{(2)}+O_0^{(3)}\}
a^\dagger_{ks}|0\rangle,
\end{equation}
where, the first term gives the creation of two electron-hole pairs
and other term gives the creation of an electron-hole pair through
the scattering of the extra electron. So that there are two types of
intermediate states, which have contributions to the correlation
energy (C1). The contribution from the intermediate state
\begin{mathletters}
\begin{equation}
|m\rangle=a^\dagger_{k_1s}a^\dagger_{k_2\bar{s}}b^\dagger_{k_3s}
b^\dagger_{k_4\bar{s}}a^\dagger_{ks}|0\rangle
\end{equation}
gives the correlation energy of the half-filled ground state and
the {\it loss} of correlation energy due to the extra electron.
The contribution from the intermediate state
\begin{equation}
|m\rangle=a^\dagger_{k_1s}a^\dagger_{k_2\bar{s}}b^\dagger_{k_3s}
|0\rangle
\end{equation}
\end{mathletters}
gives the {\it gain} of the polarization energy due to the polarized
process after the additional electron was introduced. Then we rewrite
the correlation energy in Eq.(C1) as
\begin{equation}
E_1^{\rm corr}(k)=E_g^{\rm corr}+E_\eta(k)+E_\pi(k),
\end{equation}
where, $E_g^{\rm corr}$ is the correlation energy of the ground state,
the {\it loss} of correlation energy in the ground state
\begin{mathletters}
\begin{eqnarray}
E_\eta(k_0)={U^2\over{4tN^2}}\sum_{k_1,k_2}&&{1\over
{(\epsilon_0+\epsilon_1+\epsilon_2+\epsilon_3)\epsilon_0
\epsilon_1\epsilon_2\epsilon_3}}(
\epsilon_0\epsilon_1\epsilon_2\epsilon_3+c_0c_1c_2c_3
+z_0z_1z_2z_3\nonumber\\
&&-c_0z_1z_2c_3-z_0c_1c_2z_3+2c_0c_1z_2z_3+2z_0z_1c_2c_3),
\end{eqnarray}
and the {\it gain} of the polarization energy
\begin{eqnarray}
E_\pi(k_0)=-{U^2\over{4tN^2}}\sum_{k_1,k_2}&&{1\over
{(\epsilon_1+\epsilon_2+\epsilon_3-\epsilon_0)\epsilon_0
\epsilon_1\epsilon_2\epsilon_3}}(
\epsilon_0\epsilon_1\epsilon_2\epsilon_3-c_0c_1c_2c_3
-z_0z_1z_2z_3\nonumber\\
&&+c_0z_1z_2c_3+z_0c_1c_2z_3-2c_0c_1z_2z_3-2z_0z_1c_2c_3).
\end{eqnarray}
\end{mathletters}
Here, we have defined $k_3=k_1+k_2-k_0$, $c_i=\cos k_i$,
$z_i=z\sin k_i$ and $\epsilon_i=(c_i^2+z_i^2)^{1/2}$.
So the correlation gap is
\begin{eqnarray}
E_{CG}&=&2E_\eta(\pi/2)+2E_\pi(\pi/2)\nonumber\\
&=&\kappa(z)U^2/t,
\end{eqnarray}
where
\begin{eqnarray}
\kappa(z)=\frac{1}{2N^2}\sum_{k_1,k_2}&\{&
{{\epsilon_1\epsilon_2\epsilon_3+z_1z_2z_3-c_1c_2z_3+2z_1c_2c_3}
\over{(\epsilon_1+\epsilon_2+\epsilon_3+z)\epsilon_1\epsilon_2
\epsilon_3}}\nonumber\\
&&-{{\epsilon_1\epsilon_2\epsilon_3-z_1z_2z_3+c_1c_2z_3-2z_1c_2c_3}
\over{(\epsilon_1+\epsilon_2+\epsilon_3-z)\epsilon_1\epsilon_2
\epsilon_3}}\}
\end{eqnarray}
and $k_3=k_1+k_2-\pi/2$. The coefficient $\kappa$ can be evaluated
numerically.

In the end of this Appendix, we mention that, in the second-order
perturbation theory, there is no mutual influence between the
correlated excitations and the polarized scatterings, which we
assume is small and is negligible until intermediate interaction
strength in Sec.III.

\figure{The correlation gap $E_{CG}$ for the Hubbard interaction. The
solid line is the result obtained with long-range operators
(4.3) ($d$ up to 10), the dashed line is the result with
on-site operators (4.3) ($d=0$), and the dotted line is the
perturbation result at $z=0.15$. And the dot-dashed line is
the exact correlation gap at the undimerized limit $z=0$ (Ref.\cite{lw}).
The points with error bars are the Monte Carlo result (Ref.\cite{jh}).}

\figure{The correlation gap $E_{CG}$ for the Hubbard interaction.
The dots are got by the projection technique.
The lines guide the eye. The circles are exact solutions.}

\figure{The correlation gap $E_{CG}$ for a long-range
interaction: (a) without screen $\epsilon_0=1$; (b) with
screen $\epsilon_0=3$.}

\mediumtext
\begin{table}
\setdec 0.000
\caption{The strength of electron interactions obtained by fitting to
different physical quantities.}
\begin{tabular}{ccc}
$U$ (eV) &$V$ (eV)&Physical quantities \\
\tableline
11.5& 2.4 & The groundstate energy curve $E(u)^{\rm a}$ \\
\tableline
7--9& &The dimerization $u$ and the screened force constant$^{\rm b}$ \\
\tableline
3--4& &The spin density of a neutral soliton$^{\rm c}$ \\
\tableline
3.7& 0.4&The resonant Raman spectra$^{\rm d}$ \\
\tableline
4.4& 0.8 &The band gap $E_g$$^{\rm e}$ \\
\tableline
3.7--6.0& 0.0--2.5 &The band width $W$, $u$, $E_g$ and the LO mode
$\nu_{\rm LO}$ for $N=8,10$$^{\rm f}$\\
\end{tabular}
\tablenotes{$^{\rm a}$Ref.\cite{gkgs}.}
\tablenotes{$^{\rm b}$Ref.\cite{dbkm}.}
\tablenotes{$^{\rm c}$Ref.\cite{wksk}.}
\tablenotes{$^{\rm d}$Ref.\cite{eeea}.}
\tablenotes{$^{\rm e}$Present work.}
\tablenotes{$^{\rm f}$Ref.\cite{jgea}.}
\end{table}

\end{document}